\documentstyle[12pt,aasms4]{article}

\def\deg{\ifmmode^\circ _\cdot\else$^\circ _ \cdot$\fi }    
\def\degg{\ifmmode^\circ \else$^\circ $\fi } 
\def\arcs{\ifmmode {'' }\else $'' $\fi}     
\def\arcm{\ifmmode {' }\else $' $\fi}     
\def\buildrel#1\over#2{\mathrel{\mathop{\null#2}\limits^{#1}}}
\def\mper{\ifmmode \buildrel m\over . \else $\buildrel m\over .$\fi}
\def\hper{\ifmmode \rlap.^{h}\else $\rlap{.}^h$\fi}
\def\sper{\ifmmode \rlap.^{s}\else $\rlap{.}^s$\fi}
\def\arcsper{\ifmmode \rlap.{' }\else $\rlap{.}' $\fi}
\def\arcmper{\ifmmode \rlap.{'' }\else $\rlap{.}'' $\fi}

\def\et{{\it et~al.~}}

\def\apj{ApJ}  
\def\apjl{Ap.~J.~ } 
\def\aa{Astr.~Ap. }     

\lefthead{Piccirillo et al.}
\righthead{Millimetric Observations of CMB anisotropies}

\begin{document}

\title{Millimetric ground-based observations of Cosmic Microwave Background
Anisotropy}

\author{L. Piccirillo\altaffilmark{1}, B. Femen\'\i a\altaffilmark{2},  N.
Kachwala\altaffilmark{1}, R. Rebolo\altaffilmark{2}, M. Limon\altaffilmark{1},
C.M.  Guti\'errez\altaffilmark{2}, \\ J. Nicholas\altaffilmark{1}, R. K.
Schaefer\altaffilmark{1} and  R.A.  Watson\altaffilmark{2,3}}

\altaffiltext{1}{Bartol Research Institute, University of Delaware, Newark, DE 
19716}

\altaffiltext{2}{Instituto de Astrof\'\i sica de Canarias, 38200 La Laguna,
Spain}

\altaffiltext{3}{University of Manchester, Nuffield Radio Astronomy
Laboratories, Jodrell Bank, Macclesfield, Cheshire, SK11 9DL, UK}

\begin{abstract} 

First results of a Cosmic Microwave Background (CMB) anisotropy experiment
conducted at the Observatorio del Teide (Tenerife, Spain) are presented. The
instrument is a four channel (3.1, 2.1, 1.3 and 1.1 mm) $^3$He bolometer system
coupled to a 45 cm diameter telescope. The resultant configuration is sensitive
to structures on angular scales $\sim 1\degg-2\degg$. We use the channels at
the two highest frequencies for monitoring the atmosphere, and apply a simple
method to subtract this contribution in channels 1 (3.1 mm) and 2 (2.1 mm).
The most intense structure at these two frequencies is the Galactic 
crossing
with peak amplitudes of $\sim 350$ $\mu$K. These crossings have been clearly
detected with the amplitude and shape predicted. This demonstrates that our
multifrequency observations allow an effective assessment and subtraction of
the atmospheric contribution. In the section of data at high Galactic latitude
we obtain sensitivities $\sim 40$ $\mu$K per beam. The statistical analyses
show the presence of common signals between channels 1 and 2. Assuming a simple
Gaussian auto-correlation model with a scale of coherence $\theta _c=1.32\degg$
for the signal, a likelihood analysis of this section of data reveals the
presence of fluctuations with intrinsic amplitude  $C_{0}^{1/2}
=76^{+43}_{-32}~\mu$K (68 \% CL including a $\sim$ 20\% calibration 
uncertainty). Since residual atmospheric noise might still contaminate our 
results, we also give our result as an upper limit of 118 $\mu$K at 95\% c.l.

\end{abstract}


\section{INTRODUCTION}

Measurements of fluctuations in the Cosmic Microwave Background (CMB) radiation
provide one of the most direct tests of theories for the formation of
structure in the universe. On large angular scales where the fluctuations are
produced by the Sachs-Wolfe effect in the last scattering surface, the overall
normalization $ Q_{rms-PS}\sim 18$ $\mu$K is well established by the COBE DMR
(\cite {COBE96}) and the Tenerife beam-switching (\cite{hanc96}) experiments.
On smaller scales ($\sim 1^\circ$) the acoustic effects on the last scattering
surface are expected to enhance the level of fluctuations relative to those at
larger angular scales, giving rise to a feature in the power spectrum of the
CMB fluctuations, the so-called "Doppler peak", with a position and shape which
depends on the values of cosmological parameters (the curvature of the
universe, the Hubble constant and the baryonic contribution). The current
observations have recently started to allow a first determination of the
position and height of this peak (\cite{sk96}, \cite{cat96}). Nevertheless more
observations are needed to reduce uncertainties due to residual atmospheric
signals or diffuse Galactic contamination. The main aim of the present
experiment is to fill  the  gap in $\ell$ space between the IAC-Jodrell Bank
beam-switching experiments (\cite{ten94}) with a flat spectrum weighted
$\ell=20\pm 8$, and both the ACME South Pole (\cite{sp94}) 
($\ell=68^{+38}_{-32}$) and the Saskatoon experiment (\cite{sk94}).    The
experiment consists in  a multichannel bolometer detector which is sensitive to
multipoles in the range $\ell = 38-77$ with the maximum sensitivity at
$\ell=53$. Here we present results of our first observing campaign in the
summer of 1994.

\section {INSTRUMENTAL SET-UP} 
\label{telS}

The instrument is described in detail elsewhere (\cite{lpicci,iab}).   In
summary, the optics consist of a primary off-axis parabolic mirror (45 cm
diameter) coupled to a secondary off-axis hyperbolic mirror (28 cm diameter).
The detector is a four channel photometer equipped with $^{3}$He bolometers
working at 0.33K. The bands are centered at 3.3, 2.1, 1.3  and 1.1 mm
wavelengths (channels 1, 2, 3 and 4) as defined by a combination of 
resonant  mesh filters. The instrumental noise is 3, 1, 1.6 and 1.2 mK$\cdot
\sqrt{sec}$ in thermodynamic units for channels 1 to 4 respectively. High
frequency leaks are blocked by a combination of fluorogold, black  polyethylene
and Pyrex glass filters. Laboratory tests have been done to check for the
absence of significant leaks in the filters.  The telescope is surrounded by
$45^{\circ}$  aluminum radiation shields fixed to the ground. The beam and
side-lobes have been  extensively checked by placing a distant Gunn source
oscillating within Channel 1 band. An accurate bi-dimensional map of the beam
shape has  been obtained ( 20$^\circ$ by 20$^\circ$). We also studied the far
field side lobe structures down to about $-72$ dB. These analyses show that the
beam response can be approximated by a Gaussian with FWHM=$2.^{\circ}4$ and
that no significant side-lobes are found.

The observing strategy consists of daily drift scans done at fixed position  in
azimuth and elevation. The beam throw in the sky is achieved by fixing the
secondary mirror and  chopping sinusoidally the primary mirror, according to $
\theta(t) = \theta_{0}$, $  \phi(t) = \phi_{0} + \alpha_{0}\times \sin(2 \pi
f_{w}t+ \psi)$ where $\phi (t)$ and $\theta (t)$ denote the azimuth and
elevation respectively, $(\phi_0,\theta_0) = (0^{\circ},78.5^{\circ}$) is the
initial position of the antenna (see next section), $\alpha_{0} = 2.6\degg$ is
the zero-to-peak azimuthal chopping amplitude at a reference frequency $f_{w} =
4$ Hz. The demodulation of the signal is done online in software by evaluating
the amplitude of the first (4 Hz) and second (8 Hz) harmonic of the reference
frequency. The resulting sky pattern for the transit of a point-like source
resembles respectively the well known 2-beam and 3-beam response. In this paper 
we only deal with the second-harmonic demodulated data. We measured a stable DC
offset (about 10 and 15 mK for, respectively, Ch1 and Ch2) which is partly due
to the arc-shaped motion of the beam in the sky, i.e., the axis of rotation of
the wobbling mirror is not exactly vertical, so the
center position of the beam is a few arcminutes higher than the two
lateral position. The absolute pointing  error has been estimated by observing
the millimeter emission of the full Moon;  conservatively we assumed that this
error was not larger than the diameter of the Moon ($\sim $30').  The
calibration constants, for each channel, have been chosen to give 1 K signal
when the central beam  is completely  filled with a 1 K source. The system is
calibrated using cryogenic cold loads and has been checked in the field
performing hot/cold load tests, raster scans of the Moon and of the Gunn
source. The calibrations are consistent to about 20 \% absolute accuracy.

\section {OBSERVATIONS AND DATA PROCESSING}
\label{obs}

The observations were carried out at Observatorio del Teide in Tenerife (Spain)
during June and July 1994. This observing site is at an altitude of 2400 m and
has been shown to have extremely good atmospheric transparency and stability
(Watson \et , in preparation). The precipitable water vapour during the
campaign was below 1.5 mm  for 10 \% of the time.  The observed region of the
sky was the strip at declination $\delta = 40^\circ$ where we collected about
550 hours of observations. This declination has been extensively measured from
this site  at larger angular scales ($\sim 5^\circ$) and lower frequencies (10,
15 and 33 GHz) with reported detections of structures in the CMB  by the
Jodrell Bank-IAC experiments (\cite{ten94}). 

The experiment was operated only during the night to avoid contamination from
solar radiation. Fluctuations in the atmospheric emission are  the main source
of random noise in our system.  The rms of the data collected during several
consecutive hours shows that the sky noise of the data binned to 10 s was of
the order of  5.1, 6.8, 11.5  and 13.6 mK for channels 1, 2, 3 and 4
respectively during typical observing nights. We experienced few nights with
excellent observing conditions during which the noise dropped down by a factor
$\sim$3.5 in all channels. Even in  these cases the atmospheric noise is larger
than the instrumental noise and therefore a major
goal in our processing is to assess and subtract this unwanted source of
noise.  We can  reduce  the atmospheric noise in channel 1 and 2 by subtracting
the extrapolated signal from channel 4 (1.1 mm band) which is the most
sensitive to atmospheric gradients. We checked also that extrapolating channel
3 (1.3 mm) produces similar results. In each channel $i$ we have a
superposition of astronomical signal and atmospheric signal:  $\Delta
T_{ANT,i}=\Delta T^{astro}_{ANT,i}+\Delta T_{ANT,i}^{atm} $, where all terms
are expressed in  antenna temperature ($T_{ANT,i}$) and $\Delta
T_{ANT,i}^{atm}= \alpha_{i}\,\Delta T_{ANT,4}^{atm}$.  Since the bulk of the 
signal in each channel is due to the atmospheric emission and the sky signal is
expected to be much smaller, a linear fit of channel $i$ versus channel 4
provides a very good estimation of $\alpha_{i}$.  For each channel we have to
solve the following equation to obtain the sky  signal:

\begin{equation} 
\Delta T_{ANT,i} =\frac{f_{i}}{c_{i}} \Delta T^{astro}_{i}+(\Delta
T_{ANT,4}-\frac{f_{4}}{c_{4}}  \frac{1}{\rho_{i4}} \Delta T^{astro}_{i}) \times
\alpha_{i} 
\end{equation}

where $f_{i}$ and $f_{4}$ are the atmospheric transparencies at channels $i$
and 4 computed using measurements of the water vapour content, pressure and
temperature of the atmosphere (\cite{cer85}). $\Delta T_{ANT,i}$ and $\Delta
T_{ANT,4}$ are the data in channels $i$ and 4  in antenna temperature units;
$\Delta T^{astro}_{i}$ is the astronomical signal in thermodynamic temperature
units;  $c_{i}$ and $c_{4}$ are  the Rayleigh-Jeans to thermodynamic conversion
factors:$\Delta T_{ANT,i}= \frac{1}{c_{i}} \Delta T_{i}$ ($c_i$ =1.29, 1.66, 
3.66 ,4.82 for channels 1 to 4 respectively) and $\rho_{i4}$ is the fraction of the
astronomical signals seen at channels i and 4: $\rho_{i4}=1$ for CMB signal
(i.e., outside the Galactic Plane crossing), $\rho_{i4} \neq 1$  in the area of
the Galactic  Plane crossing and evaluated according to our predictions for the
Galactic foregrounds. Notice that  the method requires a prior knowledge of the
fraction of the signal expected in the channel $i$ and in channel 4
($\rho_{i4}$). In the case of the  crossing of the Galactic Plane, we use
$\rho_{i4}$ as obtained from our estimations. However, different models in the 
literature yield essentially the same values for the ratios $\rho_{i4}$ and
threfore we are confident that the recovered  Galactic Plane is nearly model
independent.

The atmospheric cleaning procedure was run on the data binned in 10 sec 
intervals,  so that the noise is dominated by atmospheric noise in all
channels. After cleaning, the 10 s binned scans are binned again to 4 minute
bin size, so the beam is sampled with at least three points.  The typical rms
of these cleaned 4 minute binned scans are 0.33 and 0.29 mK for channels 1 and
2 respectively. These scans have an offset drift 
(always less than 0.6 mK for both channels) which we remove by
fitting and subtracting combinations of sinusoidal functions with periods equal
or larger than 72 degrees in RA, hence much larger than the scales at which our
instrument is sensitive. We then proceed with obtaining the final data set by
stacking all individual scans where the rms does not exceed 0.5 and 0.55 mK for
channels 1 and 2 respectively.  In this way we finally use 179 and 129 hours of
data for channels 1 and 2 in order to form the final data set (33\% and 23~\%
of the total data). The overlap in the data used in channel 1 and channel 2
amounts to 89 hours. Fig. 1 shows the results in thermodynamic temperatures
obtained  for channels 1 ($a$ and $b$) and 2 ($c$ and $d$) in the two sections
of data away of the Galactic plane ($|b|\stackrel{>}{_\sim} 15\degg$). The data
have been  binned in increments of 1\degg in RA. The visual appearance of the
results of channel 1 is nearly  featureless, with all the points lying below
the two-sigma level while channel 2  possibly shows the presence of signal. The
mean error-bar in the results of each channel is $\sim 70$ $\mu$K in a
1\degg~bin in RA. In Fig. 1$e$ and 1$f$ we plot the sum ((Ch 1$+$Ch2)/2) and in
1$g$ and 1$h$ the difference ((Ch  1$-$Ch2)/2) of both channels.


\section{GALACTIC FOREGROUNDS}
\label{gal}

We have analyzed the Galactic contribution including the synchrotron, free-free
and dust emission. The first two processes can be modeled by extrapolating the
low frequency surveys at 408 MHz (\cite{has82}) and 1420 MHz (\cite{reich86}).
We use a single power law to model the joint emission from these processes: 
$T_{ff-sync}\propto \nu^{-\beta}$. In the Galactic Plane crossing at
RA=$305^{\circ}$ the free-free emission is expected to dominate over synchroton
due to the presence of the source Cyg A and other unresolved HII regions and
therefore the spectral index is $\beta \sim 2.1$. This was additionally checked
by obtaining the $\beta$ values from the 1420 MHz map, which reproduce the
Galactic plane crossings at Dec=+$40^\circ$ both in the 33 GHz Tenerife
beam-switching scans (\cite{Dav96,Gut})  and in the 53 and 90 GHz COBE-DMR
maps. We obtain $\beta =2.11\pm 0.05 $, which is the value that we adopted.
With this value of $\beta$ the free-free and synchrotron contamination is
negligible at our frequencies. This is consistent with \cite{hanc96} which
found that at 33 GHz and $\delta$ = 40$^\circ$ this contamination is less than
$8\ \mu K$ on scales of about 5$^\circ$. The dust contribution was estimated by
using the 240 $\mu$m DIRBE map as Galactic template and extrapolated to our
frequencies using the model obtained in \cite{boulanger96}. According to this
model the dust emission is well described by a grey body of emissivity $\gamma$
= 2  (\cite{reach95}) and with a dust temperature $T_d$=17.5 K: $I_{\nu}\propto
\nu^{\gamma} B_{\nu}(T_d)$. In Fig 2 we show  the prediction (dashed line) for
the Galactic emission  and our results for channel 1 (top) and channel 2 
(bottom). The general agreement between the predictions and our measurements
constitute an important  check on the performance of our system and our method
of subtracting the atmosphere. Any reasonable  combination of $\gamma$ and
$T_d$ produces a negligible dust contribution outside the area of the main
Galactic Plane crossing:  $rms_{D}(ch\,1) \, = \, 0.5 \,  \mu K$ and
$rms_{D}(ch\,2) \, = \, 1.6 \, \mu K$. In this paper we have  ignored such
contribution. A more detailed analysis will be presented in a  forthcoming
paper.

\section{STATISTICAL ANALYSIS}

We have analyzed statistically the data of channels 1 and 2 in the ranges
RA$_{1}$=$241^\circ - 285^\circ$ and RA$_{2}$=$331^\circ - 20^\circ$ which are
at Galactic latitudes $|b|\stackrel{>}{_\sim}15$\degg.  We computed the
correlation function to check for the presence of common structures in both
channels. Fig. 3 shows the most relevant results obtained. In panel $a$  the
cross-correlation between channels 1 and 2 is shown, in panel $b$ we show the
auto-correlation of the combined scan (Ch 1$+$Ch 2)/2 and panel $c$ shows the
auto-correlation of the difference scan (Ch 1$-$Ch2)/2. The error-bars in each
of the three panels represent the 68\% confidence levels and have been computed
using Monte Carlo techniques. Due to the experimental configuration, the
expected correlation in our data, from signals on angular scales at which the
instrument is sensitive, will show a characteristic pattern with a positive
feature at small angles, followed by a negative bump at angles $\sim
4^{\circ}-10^{\circ}$, and then flat at larger angles. Visual evidence of a
signal with this shape exists in both $a$ and $b$ panels, indicating a 
possible correlated signal between channels. The solid lines
corresponds to the expected correlation for a model estimated using a
likelihood analysis (see below). This model has an intrinsic amplitude of 86
$\mu$K and a coherence angle $\theta _c=2.1^{\circ}$, which corresponds to the
amplitude and angle at which the maximum of the likelihood surface is attained 
for the combined analysis on channel 1 and 2.  In $c$
there is no evidence of structure, which is compatible with the
expectations in the case of pure uncorrelated noise (dashed bands).
Second, a likelihood analysis was used to determine the amplitude and origin of
the signals. Our analysis assumed a Gaussian auto-correlation function (GACF)
for the CMB signal: $C_{intr}(\theta)=C_{0} \exp(-\theta^{2}/2\theta_{c}^{2})$.
The GACF models together with its limitations and connections to more realistic
scenarios have been widely discussed in the literature (see \cite{white94}) and
we adopt it to quote preliminary results, leaving physically motivated models
for a future analysis.  Table 1 presents a summary of the likelihood results
for the coherence angle of highest sensitivity ($\theta_{c}= 1.32^{\circ}$).
Upper limits and detections are quoted at 95 \% C.L. and 68 \% C.L.
respectively. The analysis of channel 1 does not show evidence of signal in any
of the two ranges considered with a limit $C_{0}^{1/2}<72~\mu$K, while for
channel 2 we detect signal in both RA ranges. We made also a joint likelihood
analysis on channel 1 and 2 assuming that both channels have been measuring the
same signal. This gives a detection $C_{0}^{1/2} =76^{+23}_{-21}~\mu$K.
Including the error in the absolute calibration we obtain  $C_{0}^{1/2}
=76^{+43}_{-32}~\mu$K. Even though these results are compatible, we cannot
exclude a possible frequency dependence of the detected signals. A
two-dimensional (in the plane $C_0$-$\theta _c$) joint analysis of both
channels shows a well defined peak at $C_0^{1/2}$=86 $\mu$K,
$\theta _c=2.1\degg$; the expected correlation in our data  for such a signal
is plotted as the solid lines in Figs. 3$a$ and $b$.  The likelihood analysis
shows the presence of a clear signal in channel 2, while channel 1 is
compatible with noise; this indicates a source of non-CMB contaminant in
channel 2. It is unlikely that the data are contaminated significantly by the
Galaxy (see previous section).  If we convert our  results  from
($C_{0}^{1/2},\theta _c$) to band power estimates, we find $\sqrt{ \ell (\ell +
1) C_{\ell}/2 \pi} \, = \,7.7^{+11.3}_{-5.1}\cdot 10^{-10}$,  marginally
consistent with the  Standard CDM model  prediction of  $
1.7^{+0.3}_{-0.2}\cdot 10^{-10}$  at our $\overline{\ell} \, = \, 53$
(\cite{gacf_to_bp}). The excess of signal seen in Ch2 might indicate that
some residual atmospheric noise is still contaminating our results.
Therefore we also quote an upper limit of 118 $\mu$K at 95\% C.L.
A more robust atmospheric subtraction technique together with a spectral index
analysis of the measured fluctuations has been applied to our data
set confirming that part of the signal seen in Ch2 is of atmospheric origin. We
will report these new results in a future paper.

\acknowledgments

This work has been supported by a University of Delaware Research Foundation
(UDRF) grant, by the Bartol Research Institute and spanish DGICYT 
project PB 92-0434-c02 at the Instituto de 
Astrof\'{\i}sica
de Canarias. We want to thank L. Page and S. Meyer for considerable help in all
the phases of this project. A special thank for the support of L. Shulman, 
J. Poirer, R. Hoyland and the technical staff of the Observatorio del
Teide. Finally we would like to thank the staff from the Instituto Nacional 
de Meteorolog\'\i a at  Tenerife  who very kindly provided us with the 
atmospheric data used in this analysis.

\clearpage

\begin{table*}
\begin{center}
\begin{tabular}{lccc} \tableline
	
	      &RA$_{1}$ 	   & RA$_{2}$ & RA$_{1}+$RA$_{2}$\\ \tableline
 	Ch1   & $< 92 $ 	   & $<115$             & $< 72 $            \\	
	Ch2   & $94^{+63}_{-54}$   & $109^{+71}_{-64}$ 	& $101^{+50}_{-48} $ \\
  Ch1 \& Ch2  & $72^{+50}_{-45}$   & $79^{+51}_{-45} $  & $76^{+43}_{-32} $  \\
\tableline

\end{tabular}
\end{center}

\caption{Table 1. Likelihood results in $\mu$K CMB. A 20\% calibration
uncertainty has been included.}

\end{table*}

\clearpage

\section*{FIGURE CAPTIONS}

\figcaption[fig1.eps]{The stacked data sets at 3.3 mm ($a$ and $b$), 2.1 mm
($c$ and $d$), the addition ($e$ and $f$) and the difference ($g$ and
$h$). In panel 1$a$ we  show the 3-beam profile indicating the instrumental
response to a point source. Units refer to thermodynamic temperatures in this 
and following figures. \label{fig1}}

\figcaption[fig2.eps]{Comparison of data at the Galactic Plane crossing at
$l \sim 80.2 \degg$ (solid
line)  and predictions of foreground emission (dashed line) for channels 1 and
2 (see main text for details). \label{fig2}}

\figcaption[fig3.eps]{ Results of the correlation analysis: ($a$) the
cross-correlation of Ch 1 with Ch 2, ($b$)  the auto-correlation of 
(Ch1 + Ch2)/2 and ($c$) the auto-correlation of  (Ch1 - Ch2)/2. The solid lines 
in ($a$) and ($b$) is the correlation obtained with a likelihood analysis, and 
the dashed line in ($c$) is the 95 \% C.L. for uncorrelated noise. \label{fig3}}

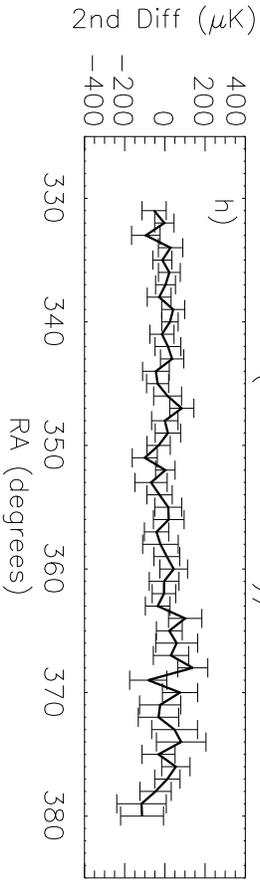
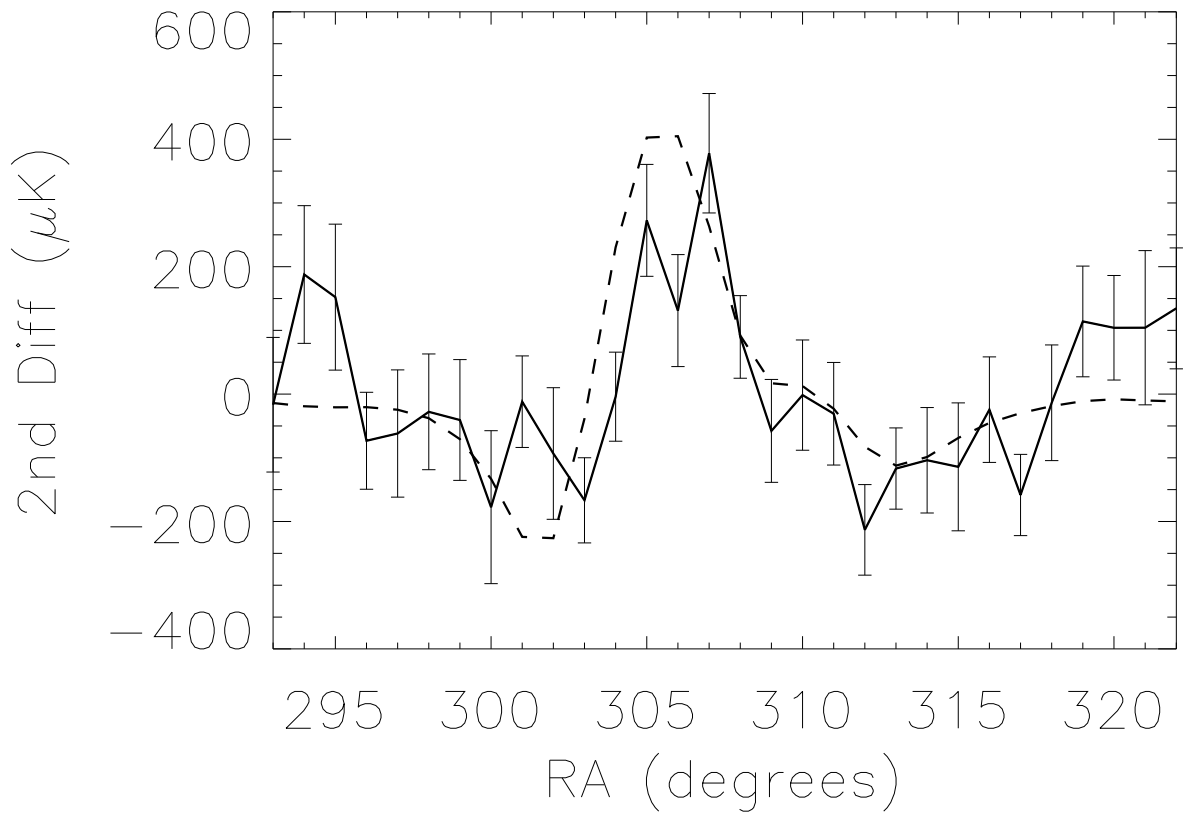
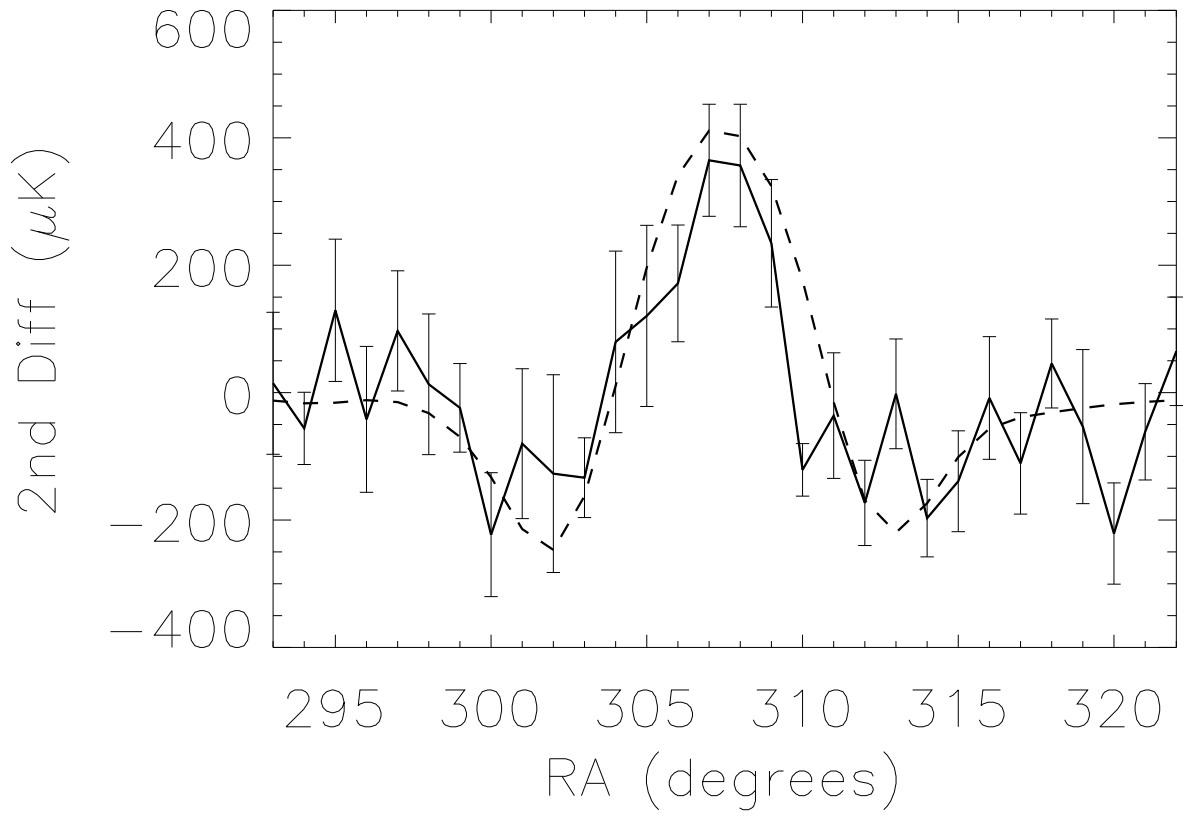
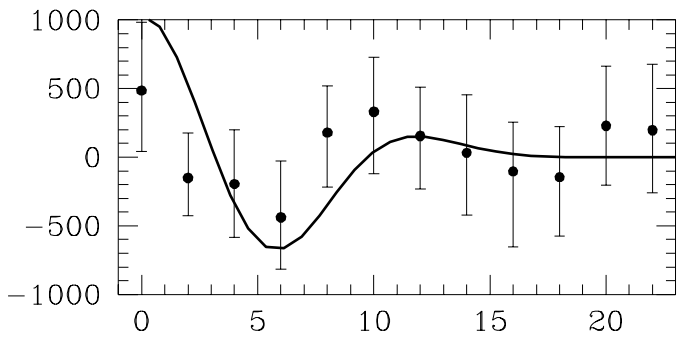
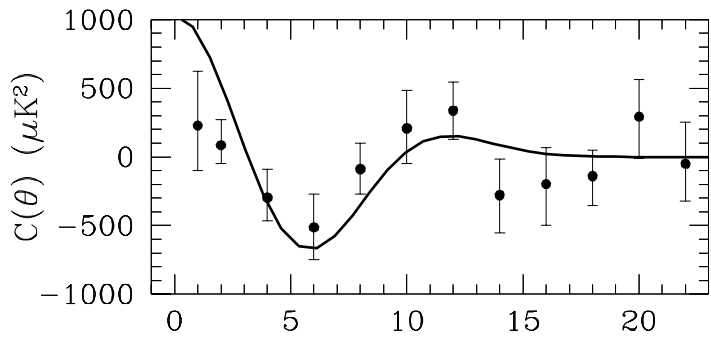
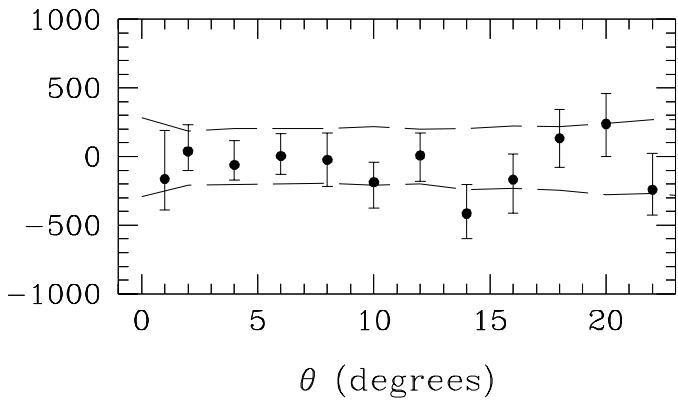
\end{document}